\documentclass[aip,rsi,preprint]{revtex4-1}

\usepackage{graphicx}
\usepackage{amsmath}
\usepackage[squaren]{SIunits}

\begin{document}
\title{Probing 
the local density of states in three dimensions with a scanning single quantum emitter}

\author{Andreas W. Schell}
\email[Electronic mail: ]{andreas.schell@physik.hu-berlin.de}
\affiliation{Nano-Optics, Institute of Physics, Humboldt-Universit\"{a}t zu
Berlin, Newtonstra{\ss}e~15, D-12489 Berlin, Germany}

\author{Philip Engel}
\affiliation{Nano-Optics, Institute of Physics, Humboldt-Universit\"{a}t zu
Berlin, Newtonstra{\ss}e~15, D-12489 Berlin, Germany}

\author{Oliver Benson}
\affiliation{Nano-Optics, Institute of Physics, Humboldt-Universit\"{a}t zu
Berlin, Newtonstra{\ss}e~15, D-12489 Berlin, Germany}

\begin{abstract}                                  
Their intrinsic properties render single quantum systems as ideal tools for quantum enhanced sensing and microscopy. 
As an additional benefit, their size is typically on an atomic scale which enables sensing with very high spatial 
resolution. Here, we report on utilizing a single nitrogen vacancy center in nanodiamond for performing three-dimensional 
scanning-probe fluorescence lifetime imaging microscopy. By measuring changes of the single emitter's lifetime information 
on the local density of optical states is acquired at the nanoscale. This technique to gather information on the local 
density of optical states is important for the understanding of fundamental
quantum optical processes as well as for the engineering of novel photonic and plasmonic devices.
\end{abstract}

\maketitle

Quantum-enhanced sensing has become one of the major fields in quantum technology~\cite{O'Brien2009}. 
The quantum properties, on the one hand, allow for measuring with a precision overcoming the 
classical limit~\cite{Jaekel1990}. On the other hand, single quantum systems represent 
ideal sensing probes. Their unique properties do not necessarily utilize intrinsic quantum effects, but 
rather the fact that quantum systems are typically very small. Therefore, they can probe the rather large 
local fields of single atoms or a local environment consisting of only a few molecules. Optical quantum 
probes provide the additional advantage of reliable initialization as well as efficient and easy read-out. 
In this paper we report for the first time on utilizing a 
single quantum emitter (QE) for performing three-dimensional scanning-probe fluorescence lifetime 
imaging microscopy (FLIM) introducing the technique of QE-FLIM.

First optical scanning quantum probes have appeared more than 10 years ago~\cite{Michaelis2000,Guthohrlein2001} 
in fluorescence microscopy. By now, recent advances with respect to the stability of solid-state emitters have 
paved the way for several novel approaches where quantum emitter scanning probes are used to detect 
electric~\cite{Dolde2011} or magnetic fields~\cite{Maze2008}.
It is also possible to detect vacuum fields via modifications of the spontaneous lifetime of an emitter,
since it is known from quantum optics that the spontaneous emission is not an intrinsic 
property of an emitter, but is determined by the local density of states (LDOS).
More generally speaking, any light-matter coupling, as described on the fundamental level by a 
single dipole coupled to modes of the electromagnetic field, can be modified by changing the LDOS~\cite{Mazzei2007}.  
By carefully designing the LDOS, it is possible to significantly enhance the 
functionality of devices in photonics and plasmonics. Examples are spontaneous emission control 
for fast optical modulators~\cite{Englund2009}, for energy-efficient lasing~\cite{Fujita2005}, or for improving light 
trapping in solar cells~\cite{Callahan2012}. The design of the LDOS is particularly crucial for fundamental 
few-photon devices in quantum optical engineering, e.g. for efficient and fast single 
photon sources needed in optical quantum computing~\cite{O'Brien2007}. Photonic structures such as microcavities~\cite{Vahala2003},  
optical antennas~\cite{Novotny2011}, and photonic metamaterials~\cite{Cortes2012} allow for a design of the LDOS in all three 
spatial dimensions. Therefore, techniques to obtain precise information about the LDOS on the 
nanoscale are crucial. There exist several approaches to gain this information, e.g. coating of 
the structures of interest with fluorescent dyes~\cite{Hoogenboom2009}, mapping with scanning near field microscopes~\cite{Imura2005,DeWilde2006},
nanopositioning of defect centers~\cite{Schell2011,Wolters2012} or colloidal quantum dots~\cite{Ropp2013},
or employing a scanning electron microscope~\cite{Sapienza2012}. Fluorescent probes indicate the LDOS via 
the observed lifetime changes, with most previous probes utilizing large ensembles of emitters
like molecules in nanobeads~\cite{Frimmer2011}.
However, because of averaging over an ensemble with different spatial positions and electromagnetic environments, 
the excitation decay curve is multi-exponential.
This makes it difficult to quantify modifications of the decay dynamics when scanning the probe.
Additionally, the ultimate spatial resolution is still given by the diameter of the doped beads.

A fundamental FLIM probe would consist of a single atom, and indeed two groups have shown the 
capability of single trapped ions to map the LDOS~\cite{Guthohrlein2001,Kreuter2004}. 
However, single ions in an ultra-high vacuum environment do not meet the requirement for a robust 
scanning-probe where a point-like fluorescent dipole is located at a scanning-tip, which can be 
actively stabilized and scanned across an arbitrary substrate. 
Defect centers in nanodiamonds~\cite{Kurtsiefer2000,Jelezko2006} provide 
optimum properties for this purpose, since they are optically stable 
even at room temperature. Their optical stability even allows for trapping with 
optical tweezers~\cite{Horowitz2012,Geiselmann2013}. 
Here, we use a single-photon emitting NV center as a scanning probe for 
fluorescence lifetime imaging. The single emitter nature of the NV center leads to 
an increase in obtainable resolution, only limited by the size of a single NV center 
and the mechanical stability of the microscope. 
With only one emitter participating, there is no temporal broadening due to lifetime variations between different emitters in
an ensemble.
Furthermore we introduce a novel mapping technique to gather three-dimensional (3D) lifetime information while keeping the
orientation of the probe fixed.
The power of this approach is demonstrated by mapping the local density of states in the vicinity of silver 
nanowires~\cite{Ditlbacher2005}.

\section*{Results}

\begin{figure*}
  \includegraphics[width=\textwidth]{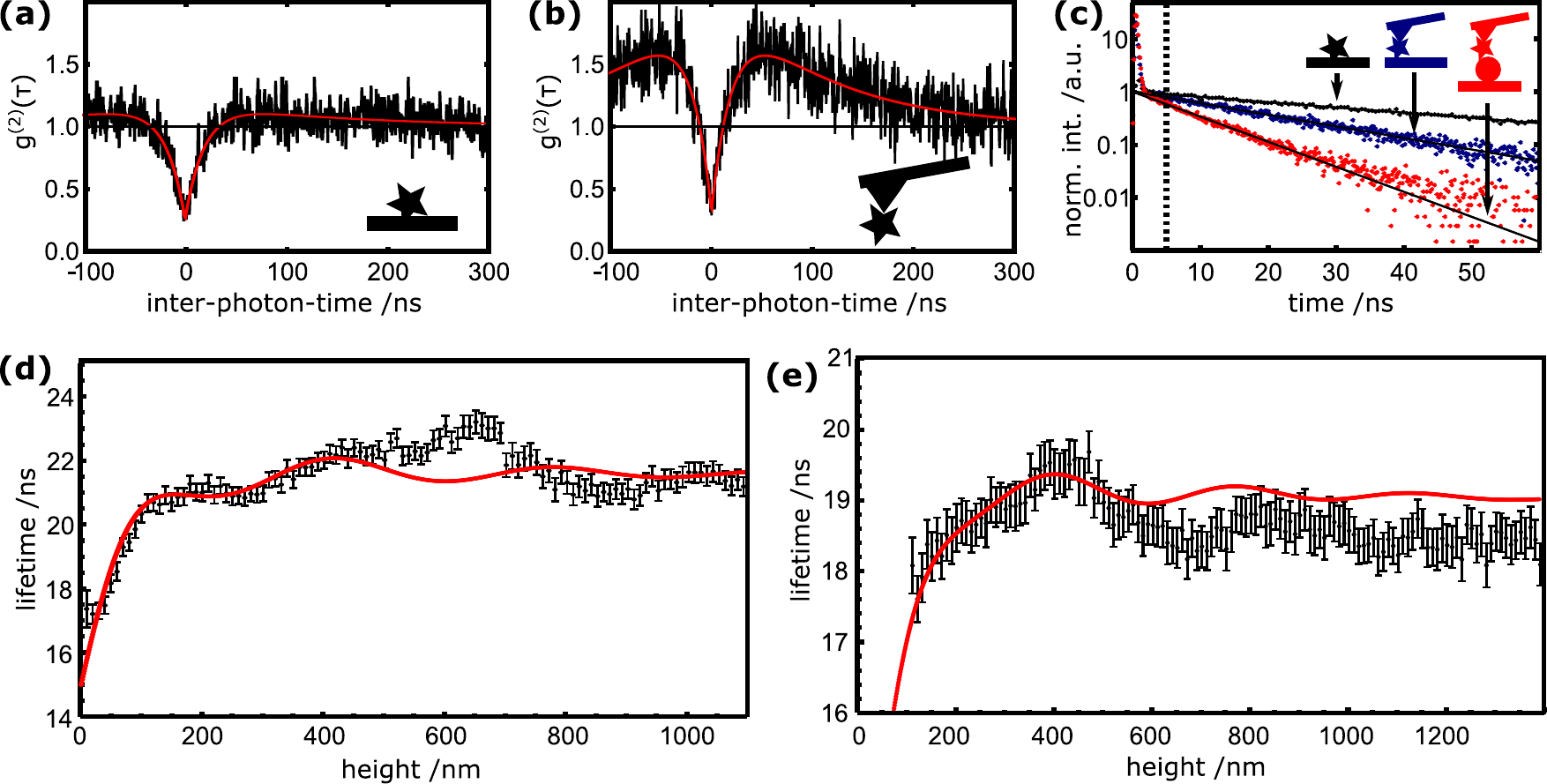}
  \caption{\textbf{Probe characterization.} \textbf{(a)} Autocorrelation function $g^{(2)}(\tau)$
  of an NV center in a nanodiamond used in the scans lying on a glass coverslip. 
    The red line is a fit to the data according to Jelezko et al.~\cite{Jelezko2006}. It yields $g^{(2)}(0)=0.25$. \textbf{(b)}
     Autocorrelation function of the NV center
  after being glued to the silicon tip.
  The red line again is a fit. It yields $g^{(2)}(0)=0.31$. Note the change in the NV center's
   bunching behavior as
  its environment changes. \textbf{(c)} Lifetime curve of the nanodiamond at the tip. The black curve is for the nanodiamond prior 
  to glueing it to the cantilever. A fit yields a lifetime of \unit{43.6}{\nano\second}. The blue and red curve are for the diamond on the tip at 
  the glass interface and at a silver nanowire, respectively. Lifetimes are \unit{19.8}{\nano\second} and \unit{9.1}{\nano\second}.
  Only photons after the dotted line are used for lifetime fits. \textbf{(d,e)} Lifetime versus height of the probe over a glass surface
  measured for two different nanodiamonds. The red curves are
  calculated according to the theory given in reference~\cite{Lukosz1977} for a dipole over
  a surface.}
  \label{fig:auto}
\end{figure*}

The experimental setup is a confocal microscope combined with an atomic force microscope for simultaneous measurements 
(see Methods and Figure~\ref{fig:setup} for details).
A very important task in quantitative QE-FLIM is the characterization of the probe. It has to be a single emitter with
known properties such as orientation, lifetime, and quantum efficiency.
To prove the single emitter character, measurements of the autocorrelation function $g^{(2)}(\tau)$ are performed before and 
after the nanodiamonds (typical sizes ca. \unit{30}{\nano\meter}) are attached to the tip (see Figure~\ref{fig:auto} (a,b)). 
A $g^{(2)}(0)<0.5$ indicates that the main photon contribution is from a single emitter~\cite{Loudon2000}. 
Only in this way averaging over
several emitters with slightly different lifetimes and slightly different positions can be excluded and 
smearing out the measurement signal is avoided.
A change in the 
bunching behavior when comparing Figures~\ref{fig:auto}(a) and (b) is attributed to a changed dielectric as well as a changed chemical
environment, which can influence the charge of the NV center~\cite{Jelezko2006,Rondin2010,Hauf2011}. 
The lifetime can be extracted from Figure~\ref{fig:auto}(c). It shows
a mono-exponential decay curve of the NV center's excited state for different situations: black for the nanodiamons on a coverslip ($\tau=46.3$~ns),
blue for the nanodiamond glued to the tip far from the surface ($\tau=19.8$~ns), and red with the tip close to a silver nanowire ($\tau=9.1$~ns).
After the nanodiamond is glued to the tip, the lifetime only varies when the environment, and therefore the LDOS, 
changes, e.g. when approaching
 a silver nanowire (red curve in Figure~\ref{fig:auto}(c)).
Figure~\ref{fig:auto} (d,e) show detailed data of nanodiamond probes approaching a glass surface.
When the emitter approaches the surface an oscillatory behavior is visible as predicted by the theory for a dipole 
over a surface~\cite{Lukosz1977}. Deviations can be explained by the fact that the silicon AFM tip
gives rise to an additional surface not covered by the theory given in Reference~\cite{Lukosz1977}. 
From the theoretical curves the NV center's orientation, its position within the nanodiamond, as well as the 
quantum efficiency of its optical transition can be derived in principle (see Supplementary Information).

With a pre-characterized and scanable single quantum emitter, the LDOS can now be mapped on the 
nanoscale in all three spatial dimensions with high resolution. In order to do this,
for each position of a scanned area the detected photons are sorted with respect to the actual height
of the oscillating cantilever (see Methods). In this way the three dimensional quantum emitter fluorescent
lifetime microscopy (QE-FLIM) is established.

\begin{figure}
  \includegraphics{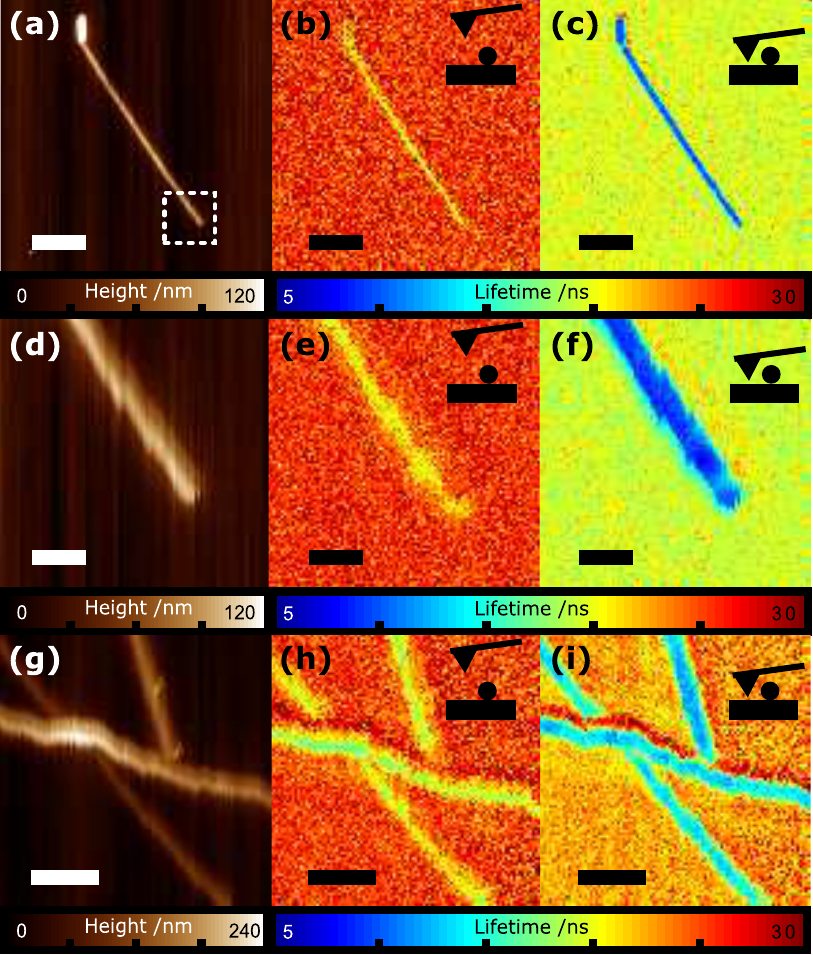}
  \caption{\textbf{Silver nanowire images by QE-FLIM.} \textbf{(a)} Atomic force microscope image of a silver nanowire of 
  diamenter \unit{100}{\nano\meter} acquired in tapping mode.
  \textbf{(b,c)} Simultaneously recorded lifetime images for the cantilever being in the most distant quarter of its oscillation and
  in the one closest to the sample, respectively. Scalebars in (a-c) are \unit{1}{\micro\meter}.
  \textbf{(d)} Zoom into the AFM image indicated by the dotted white square in (a). \textbf{(e,f)} QE-FLIM micrographs taken with the 
  nanaodiamond distant and
  close to the surface. It is apparent that the lifetime of the NV center is longer further away from the nanowire and at the same time
  its QE-FLIM image blurs out.
  Scalebars in (d-f) are \unit{200}{\nano\meter}.
  In \textbf{(g-i)} data for a network of single nanowires are shown. At 
  the nearly horizontal wire in addition to the expected decrease of the lifetime an increase is found. 
  This is is an artifact stemming from topography and the nanodiamond's position on the tip, which can be corrected (see text).
  Scalebars are \unit{500}{\nano\meter}. Cantilever oscillation amplitudes are \unit{37}{\nano\meter} for (b,c,e,f) and 
  \unit{74}{\nano\meter} in (h,i).}
  \label{fig:wires}
\end{figure}
A first example is shown in Figure~\ref{fig:wires}, where silver nanowires of approximately \unit{100}{\nano\meter} in diameter 
are raster-scanned by QE-FLIM. 
Figure~\ref{fig:wires}(a,d,g) 
are AFM topography scans showing a wire, a zoom at the wire's end, and a wire network, respectively.
In Figure~\ref{fig:wires}(b,e,h) and (c,f,i) lifetimes images
are shown for the emitter being in the distant quarter of its oscillation and in the quarter close to the surface.
The emitter's lifetime decreases close to the surface, as it is 
expected due to the higher index of refraction and therefore higher LDOS~\cite{Lukosz1977}.
Close to the nanowire the additional plasmonic
modes account for an even more reduced lifetime. Sharp features of the image of the nanowire observed when the emitter is close 
to the surface blur out when the emitter is distant.
In Figure~\ref{fig:wires}(g-i) a network of crossed wires is scanned with the QE-FLIM. Here, the cantilever's 
oscillation amplitude was set twice as large as in Figure~\ref{fig:wires}(a-f).
When scanning along the nearly horizontal wire the expected decrease of the emitter's lifetime is observed, 
but also an increase in lifetime is found in a scan parallel to it.
An explanation for this behavior is that the nanodiamond's position is not exactly at the center of the AFM tip,
so that an asymmetry arises. 
These artifacts can be avoided by taking into account topography information as 
demonstrated in the following.

\begin{figure}
  \includegraphics{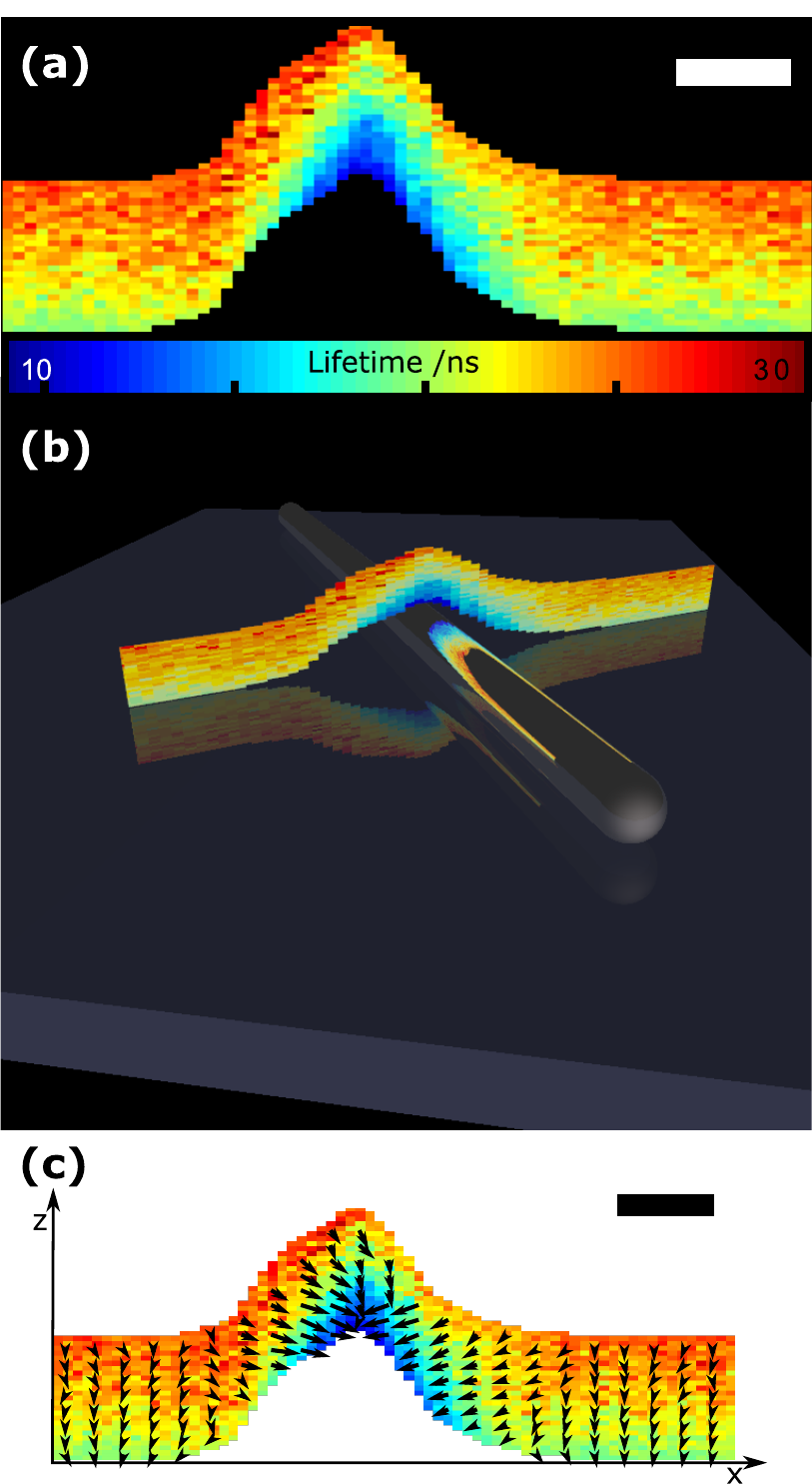}
  \caption{\textbf{X-Z QE-FLIM line scan of a silver nanowire.} \textbf{(a)} Color coded lifetime data as a function of
   height (z) and position (x). The absolute height is corrected
   according to the topography data acquired simultaneously with the AFM. In this way, artifacts are corrected as well (see text). 
   Height and position axis are scaled equally and the scalbar is \unit{100}{\nano\meter}. \textbf{(b)}
  Artists view clarifying the geometry of silver wire and data in (a). \textbf{(c)} Plot of the data in (a) with the 
  gradient of the measured lifetime in
  decay rate shown as arrows. The arrows point in the direction where the LDOS is highest, whereas the size
  of arrows indicates the absolute value of the gradient of the LDOS.}
  \label{fig:linescan}
\end{figure} 
A unique feature of 3D QE-FLIM is the possibility to derive topography corrected lifetime images. Whereas Figure~\ref{fig:wires} displayed lateral 
scans for two different relative position of the scanning cantilever, Figure~\ref{fig:linescan} shows a scan crossing a silver nanowire in a plane 
perpendicular to the sample surface. The amplitude of the cantilever's oscillation of \unit{128}{\nano\meter} is divided in 25 height bins. At the 
same time for each pixel the absolute height
of the sample is acquired with the AFM.
Therefore Figure~\ref{fig:linescan} represents a lifetime image with completely known spatial coordinates avoiding artifacts 
which often appear in scanning probe images. 
This method of topography correction can be applied to any sample. 
It is apparent 
how the lifetime decreases when approaching the glass cover slip or the surface of the silver wire. The observed rate enhancement 
lies well in the order expected from previous experiments dealing with the coupling of nanodiamonds to silver nanowires~\cite{Huck2011}. The 
highest rate enhancement found is a factor of 6.1, when comparing the diamond on the glass cover slip to the diamond coupled to 
a silver wire and a factor of for 14.5 for a gold nanosphere (see Supplementary Information). 
\begin{figure}
  \includegraphics{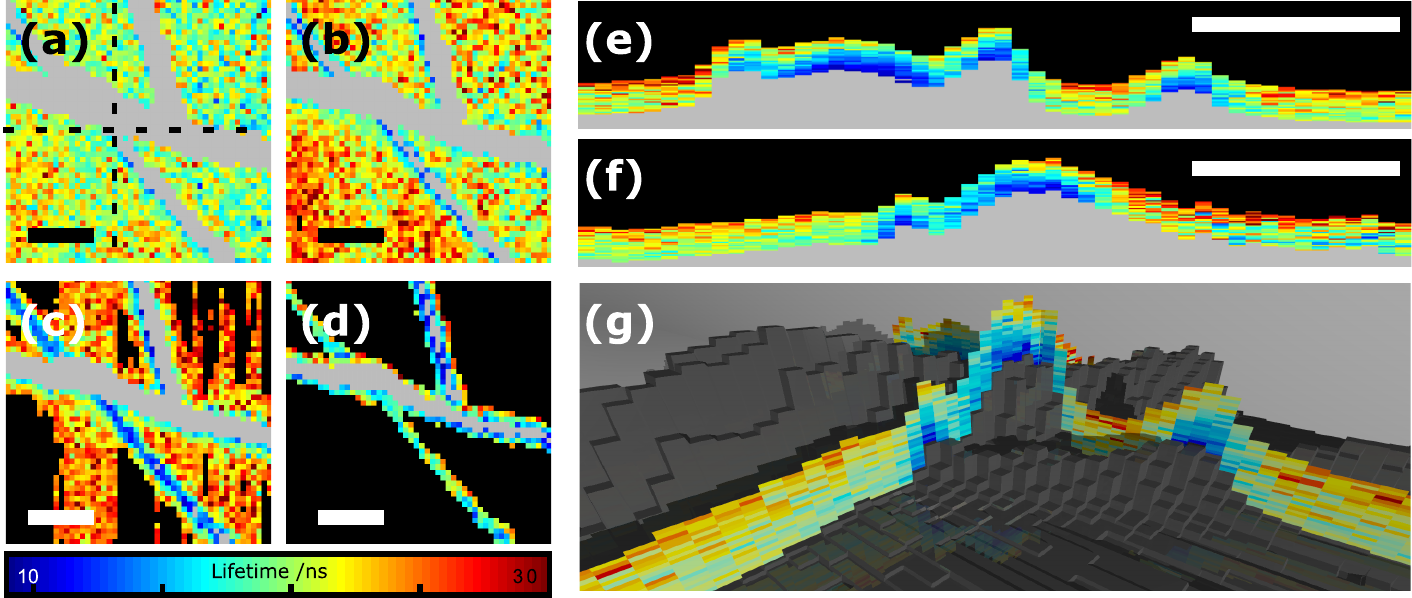}
  \caption{\textbf{QE-FLIM scans of a silver nanowires.} 
  \textbf{(a,b,c,d)} show slices through the 3D data of the nanowire network in Figure~\ref{fig:wires} at heights of \unit{53-57}{\nano\meter},
  \unit{76-80}{\nano\meter}, \unit{99-103}{\nano\meter}, and \unit{137-141}{\nano\meter}, 
  respectively. Grey areas are inaccessible 
  due to being below the surface while black areas are not reached with the set oscillation amplitude. 
  \textbf{(e,f)} show X-Z lines through the data at the positions indicated with the dashed lines in (a).
  \textbf{(g)} shows the slices from (e,f) and the AFM height data in a pseudo three dimensional manner. 
  Scalebars are \unit{500}{\nano\meter}.}
  \label{fig:corrected}
\end{figure} 
As an additional example Figure~\ref{fig:corrected}
shows lateral slices parallel to the sample with silver wires for a specific height as derived from the 
full 3D data set.  

\section*{Discussion}

From the three-dimensional topography corrected data sets it is possible to derive another important parameter, the gradient 
of the fluorescence lifetime map. Figure~\ref{fig:linescan}(c) shows a gradient QE-FLIM image of a single silver wire. 
It is apparent that the gradient 
indicates the position of nanometer-sized objects which modify the LDOS. This may be used to localize such objects even in 
topography-free sample, e.g., buried semiconductor quantum dots~\cite{Badolato2005}. As another 
important feature gradient QE-FLIM provides insight in the coupling mechanism of the quantum emitter probe and the sample. 
For example as was shown recently~\cite{J.Tisler2013} the lifetime-reduction due to Foerster-coupling depends in a characteristic
way on the dimensionality of substrate. Gradient QE-FLIM images can indicate these coupling mechanisms directly 
and with high spatial resolution.

In conclusion we have introduced a scanning probe microscope employing a single point-like quantum emitter 
as a probe for the LDOS. This technique avoids broadening of the measurement signal due to multiple emitters 
at different locations or with different lifetimes. By operation in tapping mode and collecting the photons 
tagged with the cantilever position, lifetime data in all three dimensions can be acquired in a single scan 
without topography artifacts. These three dimensional data sets give detailed insight to the electromagnetic 
environment at the nanoscale with sub-diffraction-limit resolution.

\clearpage

\section*{Methods}
\begin{figure*}
  \includegraphics[width=\textwidth]{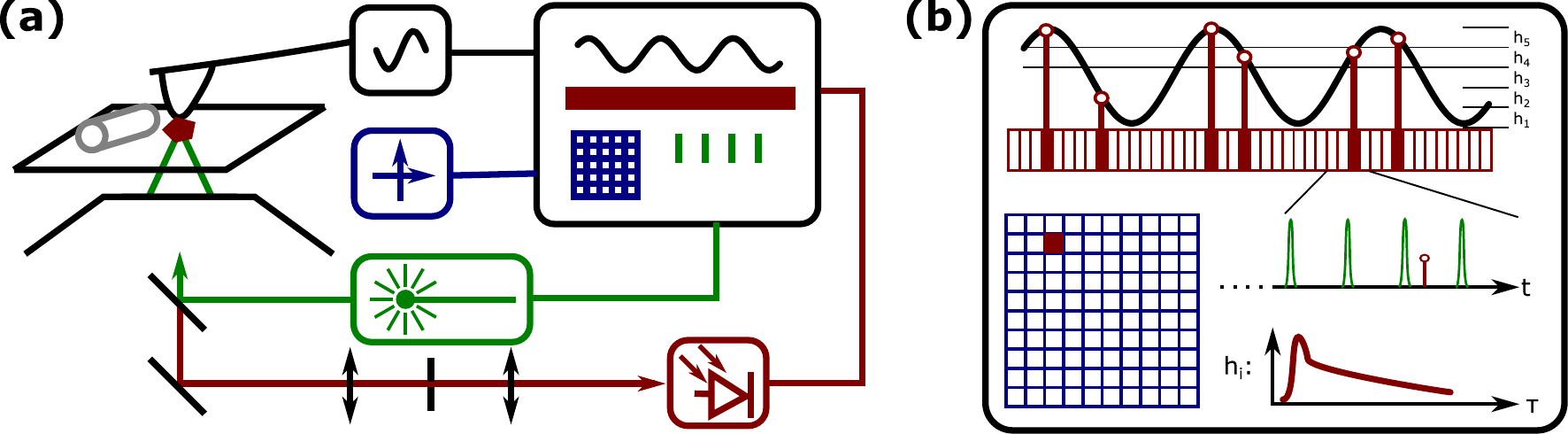}
  \caption{\textbf{Measurement scheme of QE-FLIM.} \textbf{(a)} A nanodiamond containing a single NV center is glued to the tip of an AFM cantilever.
  The AFM is operated in
  tapping mode and the cantilever is positioned in the focal volume of a sample 
  scanning confocal microscope with a pulsed laser and an avalanche photodiode.
  Cantilever oscillation, coordinates, laser timing, and photon time tags are recorded. \textbf{(b)} For each pixel 
  the photon detection events are sorted according to the 
  current height of the cantilever. The lifetime is determined for different height bins $h_i$ 
  by histogramming the arrival 
  times with respect to the laser pulse and by making a fit to the tail of the 
  exponential decay. This makes it possible to gather real 3D lifetime data.}
  \label{fig:setup}
\end{figure*}
\textbf{Confocal microscope}
The optical microscope in this experiment is a homebuilt sample-scanning inverted confocal microscope
(see Figure~\ref{fig:setup}(a)). An objective with
a numerical aperture of 1.35 (UPlanSApo60XO, Olympus) is used to excite the NV center and to collect the single photons.
Excitation is done with a picosecond laser at a wavelength of \unit{531}{\nano\meter} with a variable repetition rate 
(LDH-P-FA-530, Picoquant). A repetition rate of \unit{10}{\mega\hertz} is used when scanning with the QE-FLIM
microscope, while a rate of \unit{80}{\mega\hertz} is used for the autocorrelation measurements.
Autocorrelation measurements are performed with a Hanbury-Brown and Twiss setup.
After spectral filtering avalanche photodiodes (SPCM-AQR-14, Perkin Elmer) are used to detect
the single photons.

\textbf{Atomic force microscope}
A tip scanning atomic force microscope (NanoWizard II, JPK Instruments) is mounted on the confocal microscope
(see Figure~\ref{fig:setup}(a)). 
The nanodiamonds (MSY 0-0.05 GAF, Microdiamant AG) 
are spin coated on a cleaned coverslip, pre-characterized using the confocal microscope and subsequently attached to the
AFM cantilever. A technique similar to the one described in Reference~\cite{Schell2011b} is used 
to craft the diamonds to the tip. To make the nanodiamonds stick, the cantilevers are dipped in a drop of 
poly-L-lysine solution \cite{Cuche2009}. The typical size of the nanodiamonds glued to the tip is about \unit{30}{\nano\meter}.
Cantilevers are silicon 
tapping mode cantilevers (NSG03, NT-MDT), which proved to contribute only very little to background fluorescence.
While acquiring image data, the nanodiamond which is attached to the cantilever is positioned in the confocal volume of the microscope
and the sample is scanned with a sample scanner. 

\textbf{Electronic setup and data processing}
Photon detection event time tags are recorded using a PicoHarp 300 (Picoquant) data acquisition electronic. Markers for
spatial coordinates are also fed to the PicoHarp 300 and are collected within the time tag stream. 
To extract the 3D lifetime information, in addition to the current scanning position, 
the cantilever's oscillation is also fed to the correlation electronics.
With that information, the photon detection events at each pixel can be attributed to different height bins (see Figure~\ref{fig:setup}b). 
Time tags for the cantilevers oscillation are created from the AFM's modulation reference output
using a Schmitt trigger and additional counting electronics, so that only every 4096th period is time tagged.
Data in every pixel then is histogrammed with respect to the cantilever height and fitted with a monoexponential 
decay to calculate the lifetime (see Figure~\ref{fig:setup}(b)).
Events in the first \unit{5}{\nano\second} are not used in order to avoid artifacts from 
fast decaying background contributions. The line scan in Figure~\ref{fig:linescan} was averaged over 50
AFM scans to compensate for drift.

\section*{Acknowledgements}
Support from DFG (FOR1493) is acknowledged.

\section*{Author contributions}

A.W.S and P.E. performed the measurements and data analysis under the
supervision of O.B. A.W.S. conceived the experiment and wrote the manuscript.
All authors discussed experiment and manuscript.

%

\clearpage
\section*{Supplementary Information}

\section{Probe characterization}

For lifetime imaging with a single quantum emitter characterization of the probe is 
crucial. It is not only necessary to prove that the nanodiamond glued to the tip contains 
indeed only a single NV center,  but in addition, its quantum efficiency, position inside 
the host crystal, and orientation has also to be determined.
While it is straightforward to prove single photon emission with a Hanbury-Brown and Twiss setup,
it is more difficult to obtain information on quantum efficiency, position, and orientation.
We utilize the model of an oscillating single dipole in front of an interface~\cite{Lukosz1977_,Novotny2006_} to derive this information.
By measuring the NV center's lifetime at different heights over a coverslip 
and comparing it to the prediction by the simple model it is possible to calculate 
its quantum efficiency, position, and orientation.

The theoretical expression for the local density of optical states (LDOS) 
for a dipole in front of the interface between two media $i$ is given by
\begin{equation}
\rho(z,\lambda,\phi)=\frac{k_r}{k_{r,0}}=\frac{P}{P_0}=\cos(\phi)[\frac{P}{P_0}]_{\parallel}+\sin(\phi)[\frac{P}{P_0}]_{\perp}
\label{eq:rho}
\end{equation}
with
\begin{equation}
[\frac{P}{P_0}]_{\parallel}=\frac{3}{4}\int\limits_0^{\infty} 
Re\{\frac{s}{s_z}[r^{\perp}-s_z^2r^{\parallel}]e^{2ik_1z_0s_z}\}\,ds,
\end{equation}
\begin{equation}
[\frac{P}{P_0}]_{\perp}=\frac{3}{2}\int\limits_0^{\infty} 
Re\{\frac{s^3}{s_z} r^{\parallel} e^{2ik_1z_0s_z}\}\,ds,
\end{equation}
the Fresnel coefficients $r^{\perp}$ and $r^{\parallel}$
\begin{equation}
r^{\perp}=\frac{\epsilon_2 k_{z,1}-\epsilon_1 k_{z,2}}{\epsilon_2 k_{z,1}+\epsilon_1 k_{z,2}},
\end{equation}
\begin{equation}
r^{\parallel}=\frac{\mu_2 k_{z,1}-\mu_1 k_{z,2}}{\mu_2 k_{z,1}+\mu_1 k_{z,2}},
\end{equation}
the relations for the wave vectors $k_i$ inside medium $i$
\begin{equation}
k_{z,i}=\sqrt{k_i^2-(k_x^2+k_y^2)},
\end{equation}
and the abbreviations
$$s=\sqrt{{k_x}^2+{k_y}^2}/k_1,$$ 
$$s_z=(1-s^2)^{1/2}.$$
The LDOS $\rho(z,\lambda,\phi)$ is dependent on the dipole-interface distance $z$, the wavelength $\lambda$, and
the dipole's orientation $\phi$. It is assumed that $\mu_1=\mu_2=1$.

\begin{figure}
	\includegraphics[width=\textwidth]{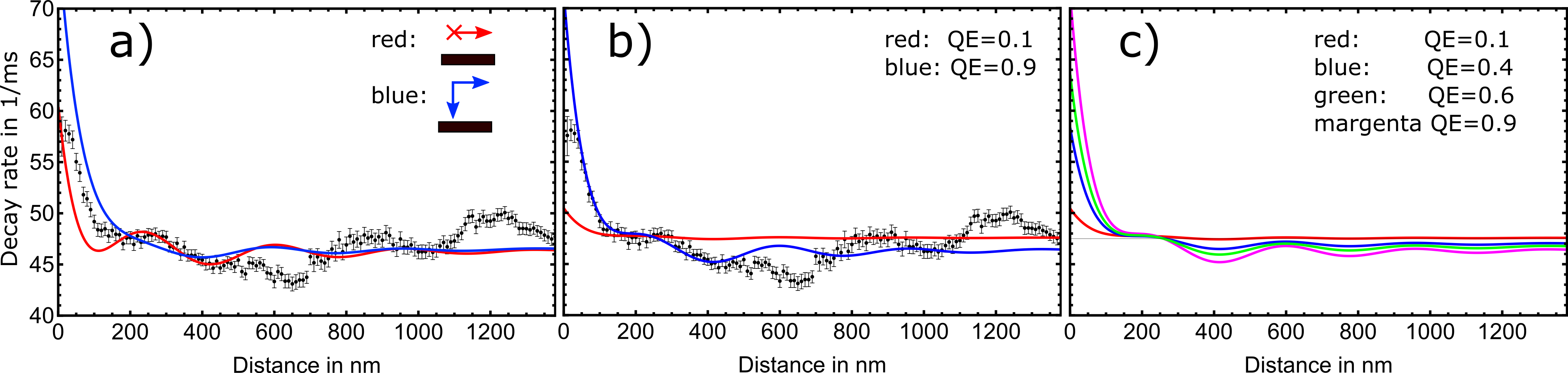}
	\caption{\textbf{Variation of different parameters in Equation~\ref{eq:rho}.} In \textbf{a} the orientation of the dipoles ($\phi$) 
	is varied. In \textbf{b} two different quantum efficiencies together with the measured data are shown. \textbf{c} shows the model
	for four different efficiencies.}
	\label{fig:params}
\end{figure}

To adapt this formula for NV centers, the LDOS for two orthogonal dipoles is added with the assumption, that both dipoles have the same 
free radiative decay rate. Figure~\ref{fig:params}a shows two extremal possibilities, both dipoles parallel to 
the surface and one parallel and one perpendicular.
Here, we approximated the broad room temperature spectrum of the NV centers 
by a Gaussian centered at \unit{700}{\nano\meter} with a standard deviation of 
\unit{50}{\nano\meter}.

The quantum efficiency QE of the NV center can now be calculated according to
\begin{equation}
QE=\frac{k_{r,0}}{k_{r,0}+k_{nr}},
\end{equation}
\begin{equation}
k(z)=k_{nr}+k_{r}=k_{nr}+k_{r,0}\,\rho(z,\lambda,\phi).
\label{kz}
\end{equation}

\begin{figure}
	\includegraphics[width=.4\textwidth]{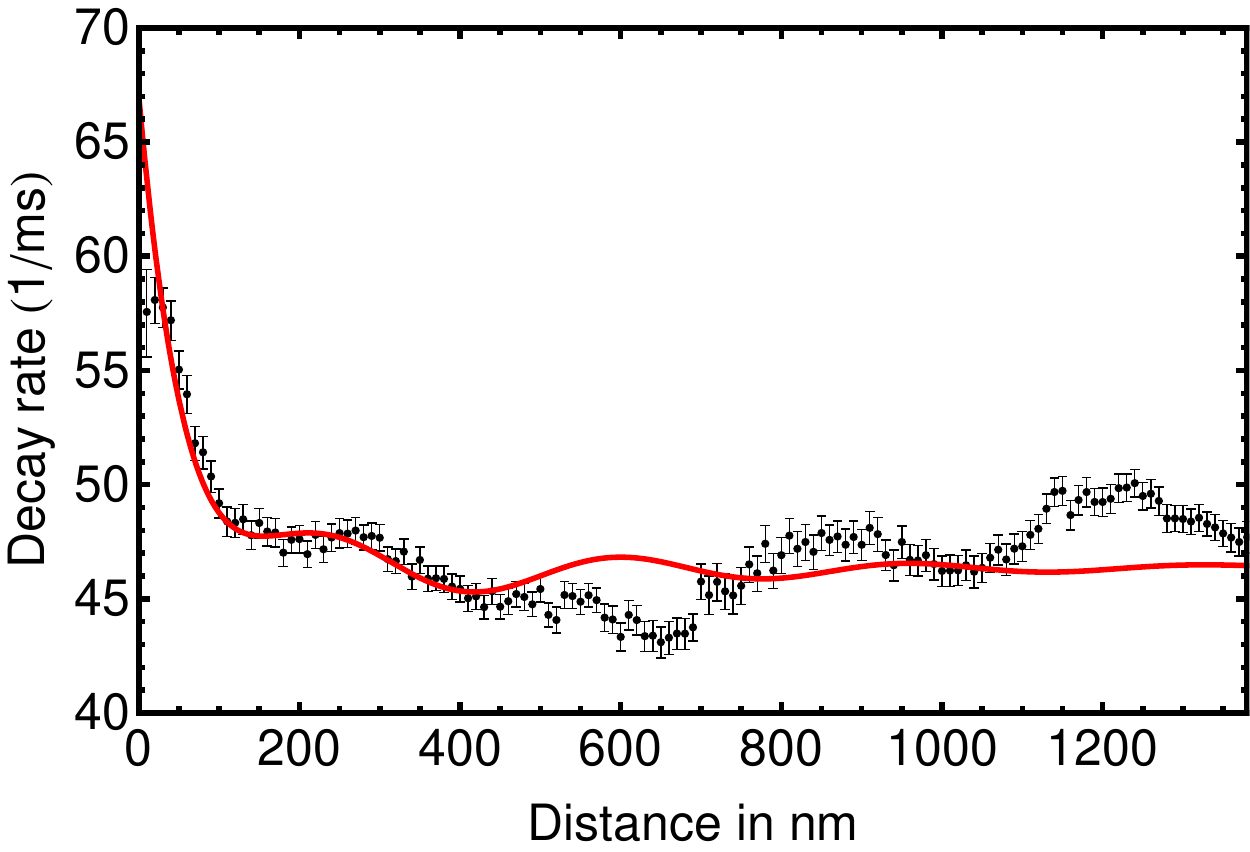}
	\caption{\textbf{Decay rate of an NV center on cantilever tip versus nanocrystal-interface distance.} The measured data are shown in black
	while the model following Equation~\ref{eq:rho} is shown in red. The values for the red curve are: 
	$k_{nr}=9~ms^{-1},~k_{r,0}=36.9~ms^{-1},~\phi=30.7$. This results in a quantum efficiency of 0.8.}
	\label{fig:data}
\end{figure}

The values of the parameters in Equation~\ref{eq:rho} now can be adjusted to fit the data measured, as shown in Figure~\ref{fig:data}.
The offset in the z axis from the position of the NV center inside the nanodiamond is set to 0, since
the quality of the data used and the accuracy of the model is not high enough to estimate values in the order of nanometers.
Deviations of data and theory are due to the fact, that the model describes a significantly simplified system -
effects arising from the interfaces of the AFM tip are not covered. Nevertheless by modeling the experimental situation in more detail,
either analytically or numerically, a better agreement might be achieved.
In the following we utilize the simpler model and show the influence of the variation of different parameters.

Figure~\ref{fig:params} shows the effect of a variation of three parameters of Equation~\ref{eq:rho}.

\section{QE-FLIM with gold spheres} 

\begin{figure}
	\includegraphics{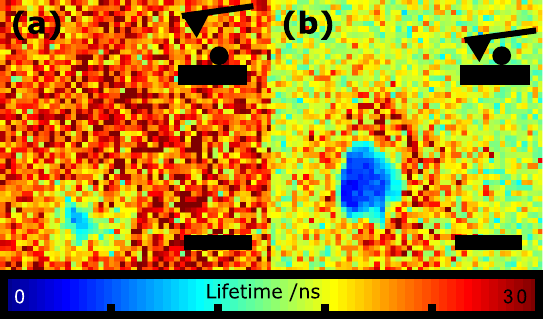}
	\caption{\textbf{A gold sphere scanned by QE-FLIM}. In \textbf{a} the NV center is  distant to the surface while in \textbf{b} it is close.
	Here the maximum enhancement by a factor of 14.5 is found. Scalebars are \unit{100}{\nano\meter}.}
	\label{fig:gold}
\end{figure}

Gold nanospheres are scanned with QE-FLIM. Figure~\ref{fig:gold} shows a scan over a gold nanosphere of \unit{90}{\nano\meter} height. 
The maximum rate enhancement found is a factor of 14.5 when the nanodiamond probe is close to the sphere's surface.

\begin{figure}
	\includegraphics[width=.7\textwidth]{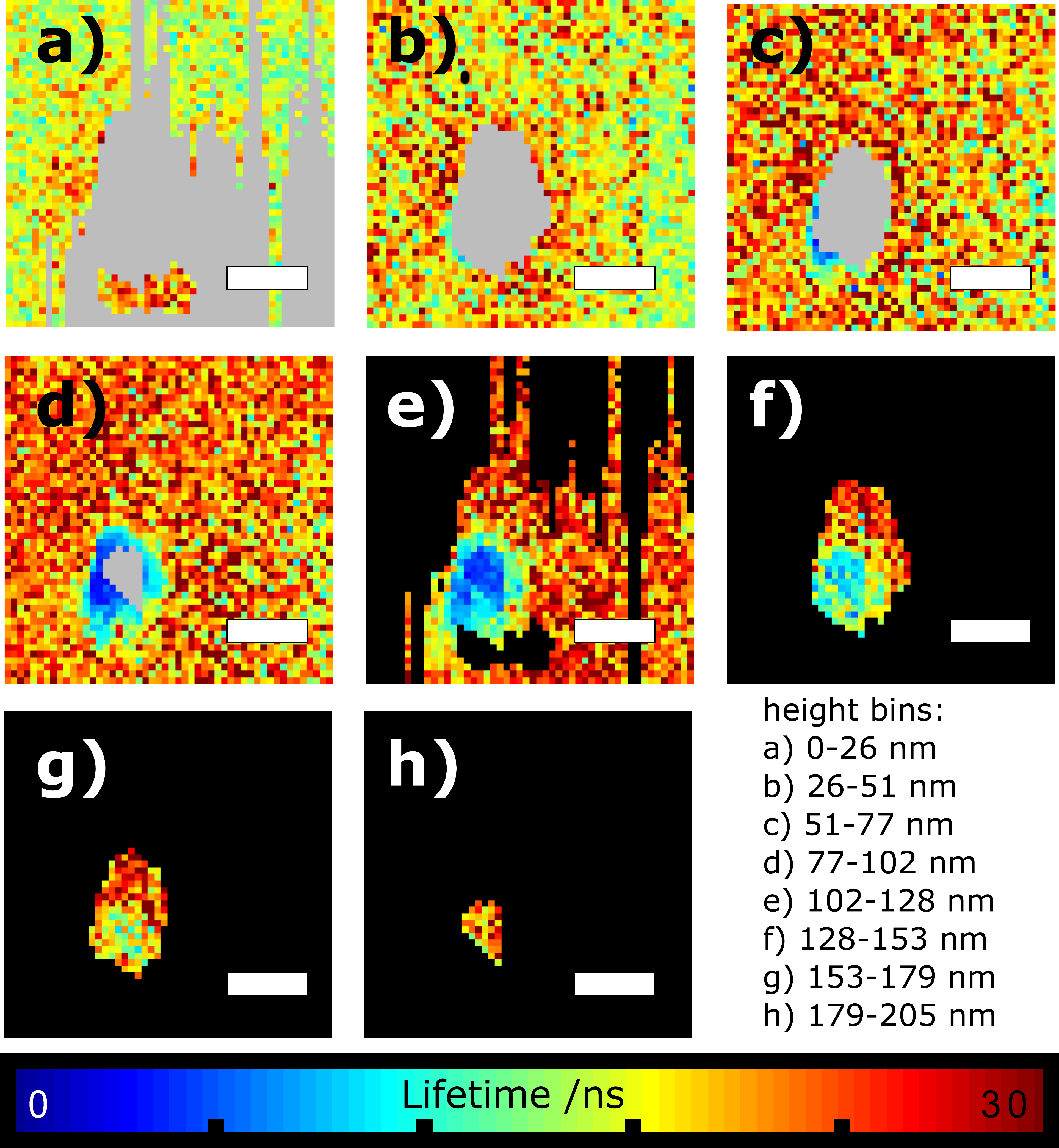}
	\caption{\textbf{A gold sphere scanned by QE-FLIM}. \textbf{(a-h)} are topography corrected images of a gold sphere scanned with QE-FLIM.
	The same data set as in Figure~\ref{fig:gold} is used.
	Scalebars are \unit{100}{\nano\meter}.}
	\label{fig:corr_topo}
\end{figure}

\end{document}